\documentstyle[12pt]{article}
\makeatletter

\long\def\@caption#1[#2]#3{\par\addcontentsline{\csname
  ext@#1\endcsname}{#1}{\protect\numberline{\csname
  the#1\endcsname}{\ignorespaces #2}}\begingroup
    \@parboxrestore
    \normalsize
    \setlength{\baselineskip}{8pt}
    \@makecaption{\csname fnum@#1\endcsname}{\ignorespaces #3}\par
  \endgroup}

\def\section{\@startsection {section}{1}{\z@}{-2.5ex plus-1ex minus
    -.2ex}{1.3ex plus.2ex}{\reset@font\large\bf}}
\def\subsection{\@startsection{subsection}{2}{\z@}{-2.25ex plus-1ex
    minus-.2ex}{0.5ex plus.2ex}{\reset@font\large\bf}}

\makeatother

\def\beq{\begin{equation}}
\def\eeq{\end{equation}}
\def\beqar{\begin{eqnarray}}
\def\eeqar{\end{eqnarray}}
\def\barr#1{\begin{array}{#1}}
\def\earr{\end{array}}
\def\bfi{\begin{figure}}
\def\efi{\end{figure}}
\def\btab{\begin{table}}
\def\etab{\end{table}}
\def\bce{\begin{center}}
\def\ece{\end{center}}

\def\text{\textstyle}

\def\si{\sigma}

\def\Ga{\Gamma}

\def\reffi#1{\mbox{Fig.~\ref{#1}}}

\def\citere#1{\mbox{Ref.~\cite{#1}}}

\def\solid{\raise.9mm\hbox{\protect\rule{1.1cm}{.2mm}}}
\def\dash{\raise.9mm\hbox{\protect\rule{2mm}{.2mm}}\hspace*{1mm}}

\newcommand{\GeV}{\unskip\,\mathrm{GeV}}

\newcommand{\TeV}{\unskip\,\mathrm{TeV}}

\def\mathswitchr#1{\relax\ifmmode{\mathrm{#1}}\else$\mathrm{#1}$\fi}

\newcommand{\PW}{\mathswitchr W}
\newcommand{\PZ}{\mathswitchr Z}

\newcommand{\PH}{\mathswitchr H}

\newcommand{\Pb}{\mathswitchr b}
\newcommand{\Pc}{\mathswitchr c}
\newcommand{\Pt}{\mathswitchr t}

\newcommand{\Zbb}{$\PZ\to\Pb\bar\Pb$}

\newcommand{\Rb}{R_\Pb}
\newcommand{\Rc}{R_\Pc}
\newcommand{\Gb}{\Ga_\Pb}

\newcommand{\Gh}{\Ga_{\mathrm h}}
\newcommand{\GT}{\Ga_{\mathrm T}}
\newcommand{\Gl}{\Ga_{\mathrm l}}

\def\mathswitch#1{\relax\ifmmode#1\else$#1$\fi}

\newcommand{\MW}{\mathswitch {M_\PW}}

\newcommand{\MZ}{\mathswitch {M_\PZ}}
\newcommand{\MH}{\mathswitch {M_\PH}}

\newcommand{\Mt}{\mathswitch {m_\Pt}}

\newcommand{\scrs}{\scriptscriptstyle}

\newcommand{\swbar}{\mathswitch {\bar s_{\scrs\PW}}}

\newcommand{\chidof}{\chi^2_{\mathrm{min}}/\mbox{d.o.f.}}

\newcommand{\alpz}{\alpha(\MZ^2)}

\newcommand{\alpsz}{\alpha_{\mathrm s}(\MZ^2)}

\newcommand{\LEP}{{\mathrm{LEP}}}
\newcommand{\SLD}{{\mathrm{SLD}}}

\newcommand{\zpC}[3]{{\sl Z. Phys.} {\bf C#1} (19#2) #3}

\newcommand{\plB}[3]{{\sl Phys. Lett.} {\bf B#1} (19#2) #3}
\newcommand{\mplA}[3]{{\sl Mod. Phys. Lett.} {\bf A#1} (19#2) #3}
\newcommand{\prD}[3]{{\sl Phys. Rev.} {\bf D#1} (19#2) #3}

\newcommand{\prl}[3]{{\sl Phys. Rev. Lett.} {\bf #1} (19#2) #3}

\def\draftdate{\relax}
\def\mda{\relax}
\def\mua{\relax}
\def\mla{\relax}
\def\draft{
\def\thtystars{******************************}
\def\sixtystars{\thtystars\thtystars}
\typeout{}
\typeout{\sixtystars**}
\typeout{* Draft mode!
         For final version remove \protect\draft\space in source file *}
\typeout{\sixtystars**}
\typeout{}
\def\draftdate{\today}
\def\mua{\marginpar[\boldmath\hfil$\uparrow$]%
                   {\boldmath$\uparrow$\hfil}%
                    \typeout{marginpar: $\uparrow$}\ignorespaces}
\def\mda{\marginpar[\boldmath\hfil$\downarrow$]%
                   {\boldmath$\downarrow$\hfil}%
                    \typeout{marginpar: $\downarrow$}\ignorespaces}
\def\mla{\marginpar[\boldmath\hfil$\rightarrow$]%
                   {\boldmath$\leftarrow $\hfil}%
                    \typeout{marginpar: $\leftrightarrow$}\ignorespaces}
\overfullrule 5pt
\oddsidemargin -15mm
\marginparwidth 29mm
}

\def\stars{\strut\leaders\hbox{*}\hfill\strut}
\def\starline{\hfil\strut\hfil\hbox to \textwidth {\stars}\hfil}

\makeatletter

\def\eqnarray{\stepcounter{equation}\let\@currentlabel=\theequation
\global\@eqnswtrue
\global\@eqcnt\z@\tabskip\@centering\let\\=\@eqncr
$$\halign to \displaywidth\bgroup\hskip\@centering
  $\displaystyle\tabskip\z@{##}$\@eqnsel&\global\@eqcnt\@ne
  \hskip 2\arraycolsep \hfil${##}$\hfil
  &\global\@eqcnt\tw@ \hskip 2\arraycolsep $\displaystyle\tabskip\z@{##}$\hfil
   \tabskip\@centering&\llap{##}\tabskip\z@\cr}
\def\appendix{\par
 \setcounter{section}{0} \setcounter{subsection}{0}
 \def\thesection{\Alph{section}}}

\makeatother

\newcommand{\lsim}
{\;\raisebox{-.3em}{$\stackrel{\displaystyle <}{\sim}$}\;}

\begin{document}

\thispagestyle{empty}
\def\thefootnote{\fnsymbol{footnote}}
\setcounter{footnote}{1}

\hfill BI-TP 96/14 \\ 
\null \hfill KA-TP-11-96 \\ 
\null \hfill May 1996 \\
\vspace*{2cm}
\begin{center}
{\large\bf IMPLICATIONS OF ELECTROWEAK PRECISION DATA \\[.2em]
ON BOUNDS ON THE HIGGS-BOSON MASS}%
\footnote{To appear in the proceedings of the XXXIth RECONTRES DE
MORIOND, ``Electroweak interactions and unified theories'', Les Arcs,
Savoie - France, March 9-16, 1996.}
\\[.8cm]
S.~Dittmaier$^1$,
D.~Schildknecht$^1$ and G.~Weiglein$^2$\\
$^1$ Fakult\"at f\"ur Physik, Universit\"at Bielefeld, D-33615
Bielefeld, Germany\\
$^2$ Institut f\"ur Theoretische Physik, Universit\"at Karlsruhe,
D-76128 Karlsruhe, Germany\\
\end{center}
\vspace*{4.0cm}
{\large\bf Abstract}\\[.2cm]
\setlength{\baselineskip}{8pt}
Carefully analyzing the dependence of the bounds on the Higgs-boson mass
$\MH$ derived from the observables $\swbar^2(\LEP)$, $\swbar^2(\SLD)$, $\Gl$, 
$\MW$, $\GT$, $\Gh$, as well as $\Rb$ and $\Rc$, and considering the 
uncertainties in $\alpz$ and $\alpsz$, we find that a stronger
bound than $40\GeV\lsim\MH\lsim 1\TeV$ at the $1\sigma$ level can hardly be 
deduced
at present, even if the experimental information on the top-quark 
mass $\Mt$ from direct production is taken into account.

%\clearpage
%\mbox{}
\clearpage
\normalsize

\vspace*{4cm}
\begin{center}
{\large\bf IMPLICATIONS OF ELECTROWEAK PRECISION DATA \\[.2em]
ON BOUNDS ON THE HIGGS-BOSON MASS}
\\[.8cm]
S.~Dittmaier$^1$,
D.~Schildknecht$^1$ and G.~Weiglein$^2$\\
$^1$ Fakult\"at f\"ur Physik, Universit\"at Bielefeld, D-33615
Bielefeld, Germany\\
$^2$ Institut f\"ur Theoretische Physik, Universit\"at Karlsruhe,
D-76128 Karlsruhe, Germany
\\[1cm]
Presented by D.~Schildknecht.
\end{center}
\vspace*{7cm}
{\large\bf Abstract}\\[.2cm]
\setlength{\baselineskip}{8pt}
Carefully analyzing the dependence of the bounds on the Higgs-boson mass
$\MH$ derived from the observables $\swbar^2(\LEP)$, $\swbar^2(\SLD)$, $\Gl$, 
$\MW$, $\GT$, $\Gh$, as well as $\Rb$ and $\Rc$, and considering the 
uncertainties in $\alpz$ and $\alpsz$, we find that a stronger
bound than $40\GeV\lsim\MH\lsim 1\TeV$ at the $1\sigma$ level can hardly be 
deduced
at present, even if the experimental information on the top-quark 
mass $\Mt$ from direct production is taken into account.
\vspace*{1.2cm}

\clearpage

\def\thefootnote{\arabic{footnote}}
\setcounter{footnote}{0}
\normalsize
\vspace{.2cm}

Various groups have recently analyzed the bounds on the mass of the
Higgs scalar which can be obtained from the precision electroweak data
in conjunction with the W-boson and the top-quark 
mass~\cite{Pass}. 
In this note we give a brief account of the results \cite{zph4} of our 
recent two-parameter $(\Mt,\MH)$ fits to the 1995 electroweak data
\cite{data8/95,SLD,MW}. 

\begin{figure}
\begin{center}
\begin{picture}(16,19)
\put(-3.0,-5.7){\includegraphics{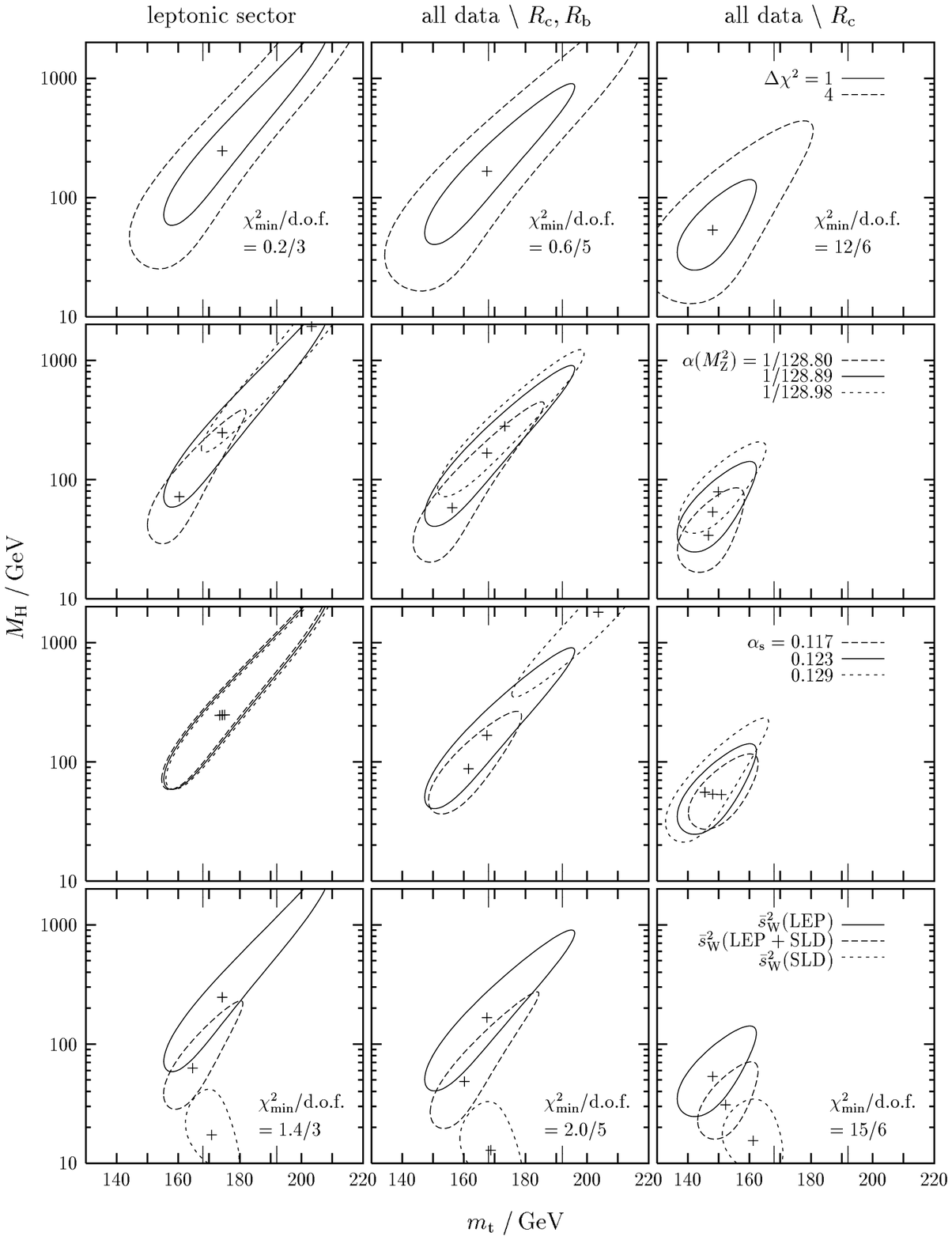}}
\end{picture}
\end{center}
\caption[xxx]{
The results of the two-parameter $(\Mt,\MH)$ fits within the SM.
The three different columns refer 
to the 
different sets of experimental data used in the corresponding 
fits, \protect\\
(i) ``leptonic sector'': $\Gl,\swbar^2(\LEP),\MW$, \protect\\
(ii) ``all data $\backslash\Rc,\Rb$'': $\GT,\Gh$ are added to set (i),
\protect\\
(iii) ``all data $\backslash\Rc$'': $\GT,\Gh,\Gb$ are added to the set (i).
\protect\\
For the fits in the first row 
the central values of 
$\alpz^{-1} = 128.89 \pm 0.09$ and $\alpsz = 0.123\pm0.006$ have been
used. The $1\si$ contours of the first row are
repeated in the other rows and compared there to
the results obtained when replacing $\alpz^{-1}$ (second row) and
$\alpsz$ (third row) by their upper and lower limits and when 
replacing $\swbar^2(\LEP)$ by $\swbar^2(\SLD)$ (fourth row). 
In all plots the empirical value of the top-quark mass,
$\Mt^{\exp} = 180 \pm 12 \GeV$, is indicated.
}
\label{fig:mtmhfit}
\efi
Our results are presented in the $(\Mt,\MH)$ plane of \reffi{fig:mtmhfit}.
The $(\Mt,\MH)$ fits have been carried out in various distinct steps 
which are indicated in the caption of \reffi{fig:mtmhfit}. This procedure 
allows us to give a detailed and transparent account of the dependence of 
the bounds on $\MH$ on the experimental input used in the fits. 

The quality of the fits in the first two columns of \reffi{fig:mtmhfit}, 
based on $(\Gl,\swbar^2,\MW)$ and $(\Gl,\swbar^2,\MW,\GT,\Gh)$,
respectively, is excellent, reflecting the well-known agreement between
Standard Model (SM) and experiment for these observables.
Based on $\swbar^2(\LEP) = 0.23186\pm0.00034$, and taking into account the 
uncertainties induced by the errors in the electromagnetic and strong 
couplings, 
displayed in the 
second and third rows of \reffi{fig:mtmhfit}, 
we find that a stronger bound than 
$40\GeV\lsim\MH\lsim 1\TeV$
can hardly be deduced from these data. Note that this conclusion remains upon
taking into account the experimental value \cite{mtexp} of $\Mt=180\pm12\GeV$. 
A significantly reduced 
future
error in $\Mt$, however, will improve the bound on $\MH$. 

If $\swbar^2(\LEP)$ is replaced by $\swbar^2(\SLD)= 0.23049\pm0.00050$
and $\swbar^2(\LEP+\SLD) = 0.23143\pm0.00028$, respectively, as shown 
in the last row of \reffi{fig:mtmhfit}, the contour in the 
$(\Mt,\MH)$ plane is strongly changed. 
In the case of 
$\swbar^2(\SLD)$, the value of $\Mt\simeq 170\GeV$ is consistent with the 
results from the direct measurements at Fermilab, 
while the values for 
$\MH=17^{+25}_{-9}\GeV$ and $\MH=13^{+20}_{-7}\GeV$ 
are in conflict with 
the lower bound \cite{MH} of $\MH>65.2\GeV$ from the Higgs-boson searches at 
LEP. With the SLD result for $\swbar^2$ taken by itself, the Higgs 
mechanism of the unmodified 
SM is in serious trouble. 
It is the prevailing opinion at present that the 
$2\sigma$ shift between $\swbar^2(\LEP)$ and $\swbar^2(\SLD)$ is due
to statistical fluctuations. Accordingly, upon using the average of 
$\swbar^2(\LEP+\SLD)$ in the fits, one obtains 
relatively low best-fit values
for $\MH$ which are consistent with the lower bound of $\MH>65.2\GeV$.
Note that the corresponding upper $1\si$ bound for $\MH$ is similarly
sensitive against variations in $\alpz$ and $\alpsz$ (not shown in the
plot) as the one obtained by using $\swbar^2(\LEP)$ (as shown in the
plot).

According to the last column of \reffi{fig:mtmhfit}, a drastic shift of 
the contour in the 
$(\Mt,\MH)$ plane towards low values of $\Mt$ and $\MH$ occurs when including 
$\Rb$ in the set of observables being fitted. At the same time the quality of 
the fit changes drastically to $\chidof = 12/6$ in the last column,
which signals the well-known discrepancy between SM prediction and
experimental result for $\Rb$,
and the $1\si$ bounds on $\MH$ are much tighter. 
Moreover, the sensitivity against variations of $\alpz^{-1}$ and
$\alpsz$ is considerably weaker.

\begin{figure}
\begin{picture}(17,7.5)
\put(3.5,-15.8){\includegraphics{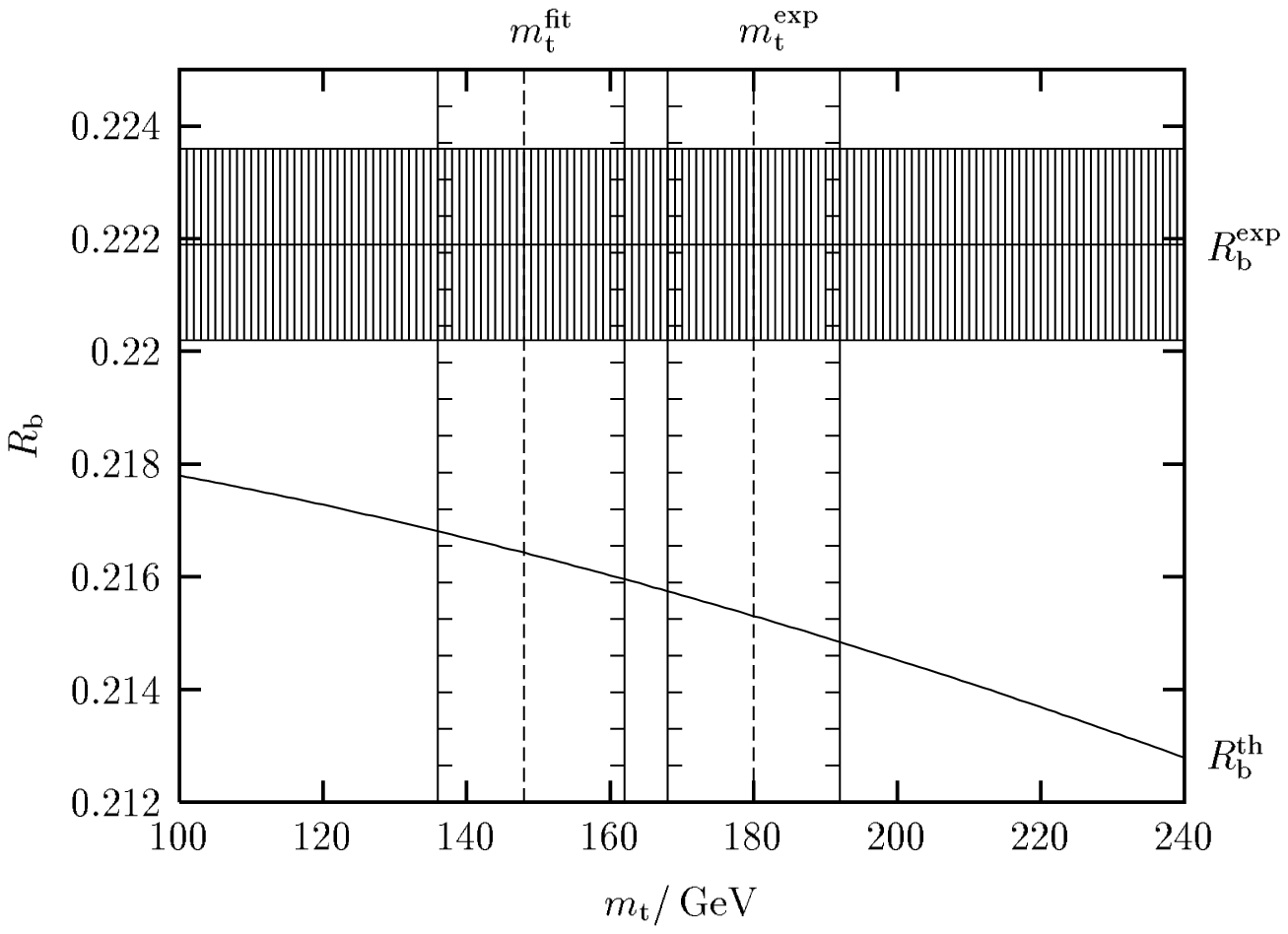}}
\put(0.0,4.0){
\parbox{5.5cm}{ \caption{
The theoretical value of $\Rb$ as a function of $\Mt$ is compared to 
$\Rb^{\exp}$. The sensitivity of $\Rb$ on $\MH$ is so small that it 
is invisible in the plot. The fit result for $\Mt$,
$\Mt^{\mathrm{fit}}$, obtained in the $(\Mt,\MH)$ fit shown in the third
column of \reffi{fig:mtmhfit},
and the Fermilab experimental value, $\Mt^{\exp}$, are also indicated.  
\label{fig:Rb}
} } }
\end{picture}
\efi
A look at \reffi{fig:Rb} is illuminating in order to understand the 
results in the last column of \reffi{fig:mtmhfit}. 
The SM prediction $\Rb^{\mathrm{th}}$ for $\Rb$ increases appreciably with
decreasing values of $\Mt$, but is extremely insensitive against
variations in $\MH$.
Inclusion of the (enhanced) experimental value of $\Rb^{\exp}$ in the 
fit
necessarily leads to a strong decrease in $\Mt$, viz.\
$\Mt^{\mathrm{fit}} = 148^{+14}_{-12}\GeV$, a value which is 
significantly below the experimental value of $\Mt=180\pm12\GeV$. 
The $(\Mt,\MH)$ correlation in the 
contours corresponding to the set of observables 
$(\Gl,\swbar^2,\MW,\GT,\Gh)$ then yields the low values of $\MH$ which 
appear in the last column of \reffi{fig:mtmhfit}. 
Effectively, including $\Rb$ in the fit thus 
appears equivalent to imposing a fixed (and low) value of $\Mt$. 
In fact, chosing a low value such as $\Mt=148\GeV$ as input in a
single-parameter $\MH$ fit to $(\Gl,\swbar^2,\MW,\GT,\Gh)$, one arrives
at values of $\MH<100\GeV$ (see discussion in \citere{zph4}).
Moreover,
when evaluating $\Rb$ for the best-fit values (using 
$\swbar^2(\LEP)$) of $(\Mt,\MH)=(148^{+14}_{-12}\GeV,54^{+88}_{-29}\GeV)$,
the resulting theoretical prediction,
$\Rb^{\mathrm{th}} = 0.2164^{-0.0005}_{+0.0004}$
(with the errors indicating the changes by varying $\Mt$ within the $1\si$ 
limits), still lies more than $3\sigma$ below
the experimental value of $\Rb = 0.2219 \pm 0.0017$.
In addition, the low value for $\Mt^{\mathrm{fit}}$ is at variance with
the Fermilab measurements so that the 
low central value and tight bound obtained for $\MH$ by including
$\Rb$ in the fit appears unreliable.
This conclusion is strengthened by the fact that a simple
phenomenological modification of the \Zbb\ vertex, as discussed 
in \citere{zph4}, leads to values of $\Mt$ compatible with the result of
the direct searches and removes the stringent upper bounds on $\MH$.

In \citere{zph4} also the effect of including/excluding the observable
$\Rc$ has been analyzed. The results for $\Mt$ and $\MH$ are only very
weakly affected by including $\Rc$. 

In summary, we find that the 
SM fits to the precision data at the 
Z-boson resonance and $\MW$ favor a Higgs-boson mass lying in the 
perturbative regime of $\MH\lsim 1\TeV$. Keeping in mind the sensitive 
dependence of the fit results for $\MH$ on the data on $\Rb$ and 
$\swbar^2(\SLD)$ and on variations in $\alpz$ and $\alpsz$,
we conclude that a stronger upper bound on $\MH$ can hardly be 
deduced from the data at present.

\end{document}